\documentclass{iopart}
 
\usepackage{epsfig}
\usepackage{here}
\usepackage{cite}

\begin{document}
   \title[]{Two-proton correlation function for the $pp \to pp+\eta$ and $pp\to pp+pions$ reactions
   }
\author{P.~Klaja$^{1,2,3}$\footnote{e-mail address: p.klaja@fz-juelich.de}, 
        P.~Moskal$^{2,3}$\footnote{e-mail address: p.moskal@fz-juelich.de},
        E.~Czerwi\'nski$^{2,3}$, R.~Czy{\.z}ykiewicz$^3$, A.~Deloff$^4$, D.~Gil$^3$, D.~Grzonka$^2$,
        B.~Kamys$^3$, A.~Khoukaz$^5$, J.~Klaja$^{2,3}$, W.~Krzemie\'n$^{2,3}$,
        W.~Oelert$^2$, J.~Ritman$^2$, T.~Sefzick$^2$,
        M.~Siemaszko$^6$, M.~Silarski$^3$, J.~Smyrski$^3$, A.~T{\"a}schner$^5$, 
        M.~Wolke$^2$, J.~Zdebik$^3$, M.~Zieli\'nski$^{2,3}$, W.~Zipper$^6$
      }
\address{ $^1$ Physikalisches Institut, Universit{\"a}t Erlangen--N{\"u}rnberg, D-91058 Erlangen, Germany}
\address{ $^2$ Institut f{\"u}r Kernphysik, Forschungszentrum J\"{u}lich, D-52425 J\"ulich, Germany}
\address{ $^3$ The Marian Smoluchowski Institute of Physics, Jagellonian University, PL-30-059 Cracow, Poland}
\address{ $^4$ The Andrzej Soltan Institute for Nuclear Studies, PL-00-681 Warsaw, Poland}
\address{ $^5$ Institut f{\"u}r Kernphysik, Westf{\"a}lische Wilhelms--Universit{\"a}t,  D-48149 M{\"u}nster, Germany}
\address{ $^6$ Institute of Physics, University of Silesia, PL-40-007 Katowice, Poland}

\begin{abstract}
For the very first time, the correlation femtoscopy method is applied
to a kinematically complete measurement of meson production in the collisions of hadrons.
A two-proton correlation 
function was derived from the data 
for the $pp\to ppX$ reaction, measured near the threshold of $\eta$ meson production. 
A technique developed for the purpose of this analysis
permitted to establish the correlation function
separately for the production of the $pp+\eta$
and of the $pp+pions$ systems.
The shape of the two-proton correlation function
for the $pp\eta$ differs from that for the $pp(pions)$ and both
do not show a peak structure opposite to   
results determined for inclusive measurements
of heavy ion collisions.  
\end{abstract}

 
\submitto{\JPG}
 
\section{Introduction}
Momentum correlations of particles at small relative velocities are
widely used to study the spatio-temporal characteristics of
 the production processes
in relativistic heavy ion collisions~\cite{lisa}.
This technique, called after Lednicky {\em correlation femtoscopy}~\cite{led}, 
originates from photon intensity interferometry
initiated by Hanbury Brown and Twiss~\cite{hbt}.
Implemented to nuclear physics~\cite{led,koonin,kopyl}
it permits to
determine the duration of the emission process and the size
 of the source from which the particles are emitted~\cite{led}.
A central role plays the correlation function
which has been defined as the measured
two-particle distribution normalized to a reference spectrum obtained by mixing particles
 from different events~\cite{led}.
 The importance of the correlation femtoscopy
 has been well established for investigations of the dynamics of
heavy ion collisions with high multiplicity.
However, as pointed out by Chajecki~\cite{chaj},
in the case of low-multiplicity collisions
the interpretation of correlation function
measurements is still not fully satisfactory,
especially in view of the 
surprising observation by the STAR
collaboration 
 indicating universality of the resulting femtoscopic radii
for both,
the hadronic (proton-proton), and heavy ion collisions~\cite{chajlis}.
One of the  challenging issues in this context 
is the understanding of contributions 
from non-femtoscopic correlations
which may be induced by the decays of resonances,
global conservation laws~\cite{chaj},
or by other unaccounted interactions.
 In contrast to heavy ion collisions, in the case of single meson production,
 the kinematics of all ejectiles may be entirely determined and hence
 a kinematically complete measurement of meson production
 in the collisions of hadrons
 gives access to complementary information
 which could shed light on the interpretation of the two-proton 
 correlations observed in heavy ion reactions.
 It is also important to underline that the 
 correlation of protons was never exploited till now in near threshold
 meson productions, and as an observable
 different from the distributions of cross sections, 
 it may deepen our understanding
 of the dynamics of meson production.
 Particularly favourable are exclusive experiments conducted
 close to the kinematical threshold where
 the fraction of the available phase-space associated
 with low relative momenta between the ejectiles is large~\cite{review}.
\par
 In this article we report on a $\eta$ meson and multi-pion production experiment
 in which the mesons were created
 in collisions of protons at a beam momentum of 2.0259~GeV/c corresponding 
 to an excess energy of Q = 15.5~MeV for the $pp\to pp\eta$ reaction.
 The measurement of the two-proton correlation function for these reactions
 is  important not only in the context of 
 studying the dynamics underlying the heavy ion physics.
 Such investigations are interesting by themselves
 because they offer a new promising diagnostic tool,
 still not exploited, for studying the dynamics of meson production
 in hadron collisions.
\par
 The correlation function carries information about the
 emitting source and, in particular,
 about the size of the interaction volume of
 the $pp\to pp\eta$ process. The knowledge of this size
 might be essential to answer the intriguing question whether
 the three-body $pp\eta$ system is capable of supporting
 an unstable Borromean bound state.
 Borromean systems may be realized in a variety
 of objects on the macroscopic (e.g. strips of papers),
 molecular~\cite{chichak, cantrill}
 and nuclear
 scale (e.g. $^{11}$Li or $^6$He nuclei~\cite{zhukov, marques, bertulani}).
 According to Wycech~\cite{wycech_acta},  the large enhancement
 of the excitation function for the $pp\to pp\eta$ reaction observed
 close to the kinematical threshold
 may be explained by assuming that the proton-proton pair is emitted
 from a large (Borromean like) object whose radius is about 4~fm.

\section{Experimental technique}
\begin{figure}[h]
\begin{center}
\includegraphics[width=7.0cm]{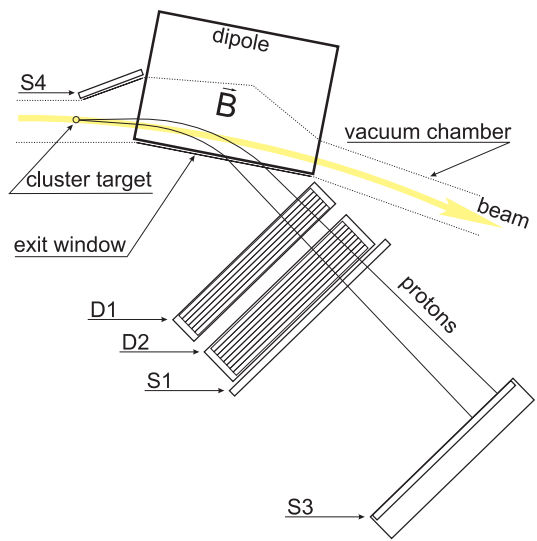}
\end{center}
\caption{Schematic view of the COSY-11 detection setup~\cite{brauksiepe}. 
Superimposed lines indicate trajectories of  protons from the $pp\to pp\eta$ reaction.
The positively charged particles were identified using the 
time of flight between the scintillator detectors S1 and S3 combined with a 
momentum reconstruction by tracking back the trajectories (reconstructed based on the signals from the  
drift chambers D1 and D2) through the magnetic field of the COSY dipole to 
the reaction point. The size of the detectors and their relative 
distances are not to scale. More detailed description can be found 
in~\cite{brauksiepe,habil,etajurek,prc69}.}
\label{fig:cosy}
\end{figure}

The experiment was conducted using the proton beam of the cooler synchrotron COSY~\cite{cosy}
and an internal hydrogen cluster target~\cite{domb}.
Momentum vectors of outgoing protons from the $pp\to ppX$ reaction
were measured by means of the COSY-11 facility~\cite{brauksiepe} presented schematically
in Figure~\ref{fig:cosy}.

The two-proton correlation function $R(k)$
was determined for the $pp\eta$ and $pp(pions)$ systems, respectively.
Here, $R(k)$ denotes
a projection of the correlation function
onto the momentum of one of the protons in the proton-proton center-of-mass 
system\footnote{
Note, that some authors instead of $k$ take  as the independent variable
the 
relative momentum of emitted
particles $q = |\vec{p_{1}}-\vec{p_{2}}|$ with $k \approx q/2$.}.
It was calculated, by means of the well established technique,
as a ratio of the momentum ($k$) dependent reaction yield
$Y(k)$ to the uncorrelated
yield $Y^*(k)$ according to the formula (cf.~\cite{boal})
\begin{equation}
   R(k)+1 = C~\frac{Y(k)}{Y^*(k)},
\label{equ:3}
\end{equation}
where $C$ denotes an appropriate normalization constant.
$Y^*(k)$ was derived from the uncorrelated reference sample obtained
 by using the event mixing technique introduced by Kopylov and Podgoretsky~\cite{kopyl}.

The separation of the correlations of the $pp+\eta$ from the $pp+pions$
system and corrections for the limited acceptance 
of the detection system constitute 
the two main challenges to be solved when deriving the correlation function
from the experimental data.

\subsection{Separation of events from the production of $pp\eta$ and $pp+pions$ systems}
In the discussed experiment, the four-momenta of the two final state protons were measured only, and the unobserved meson
was identified via the missing mass technique~\cite{habil,prc69}.
In such situations the entirely accessible information about the reaction is contained
in the momentum vectors of the registered protons.
Therefore, it is impossible to know
whether in a given event the $\eta$ meson or a few pions have been created.
However, statistically one can separate these groups of events
using the missing mass spectra 
for each  chosen region of the phase-space.

\begin{figure}[h]
\includegraphics[width=.32\textwidth]{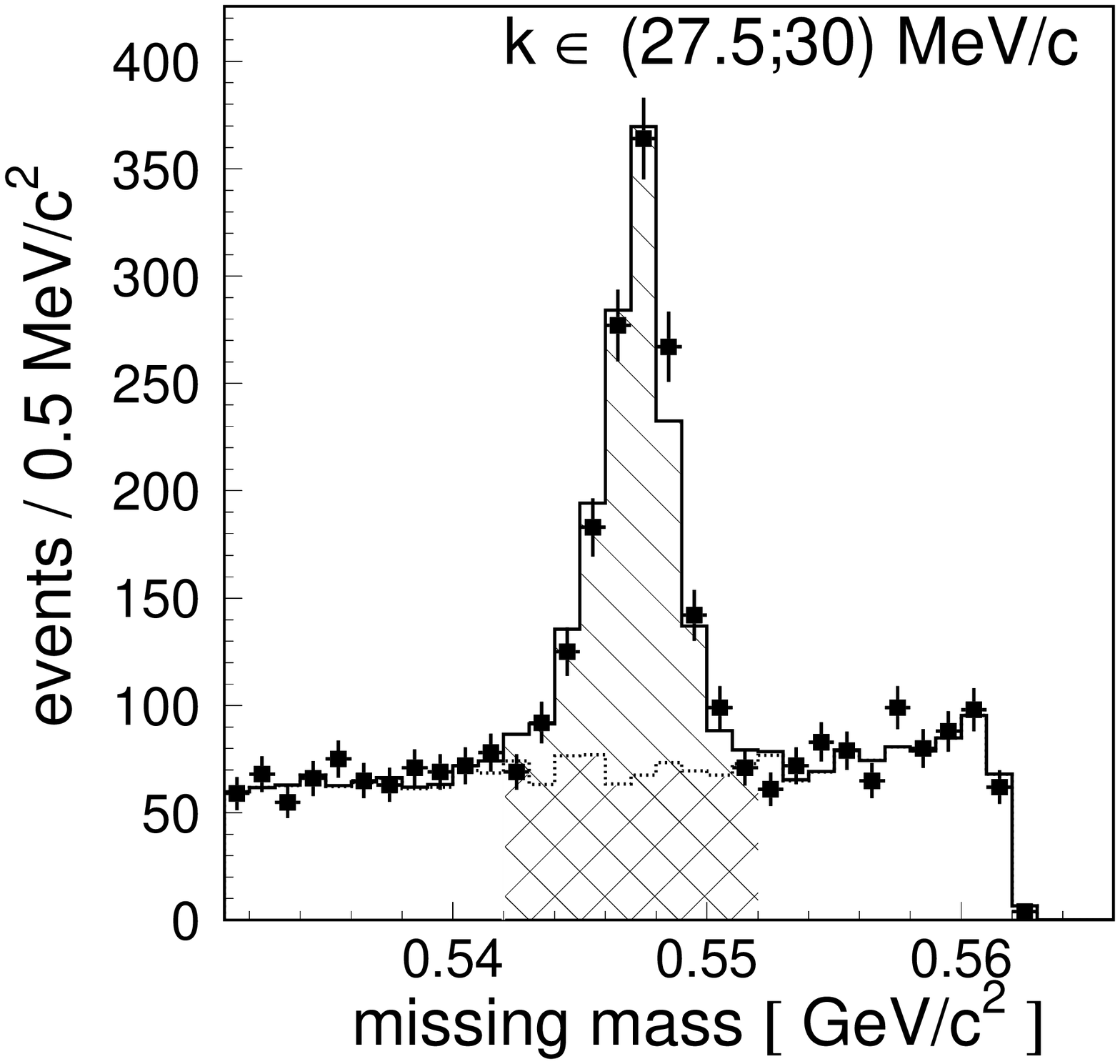}
\includegraphics[width=.32\textwidth]{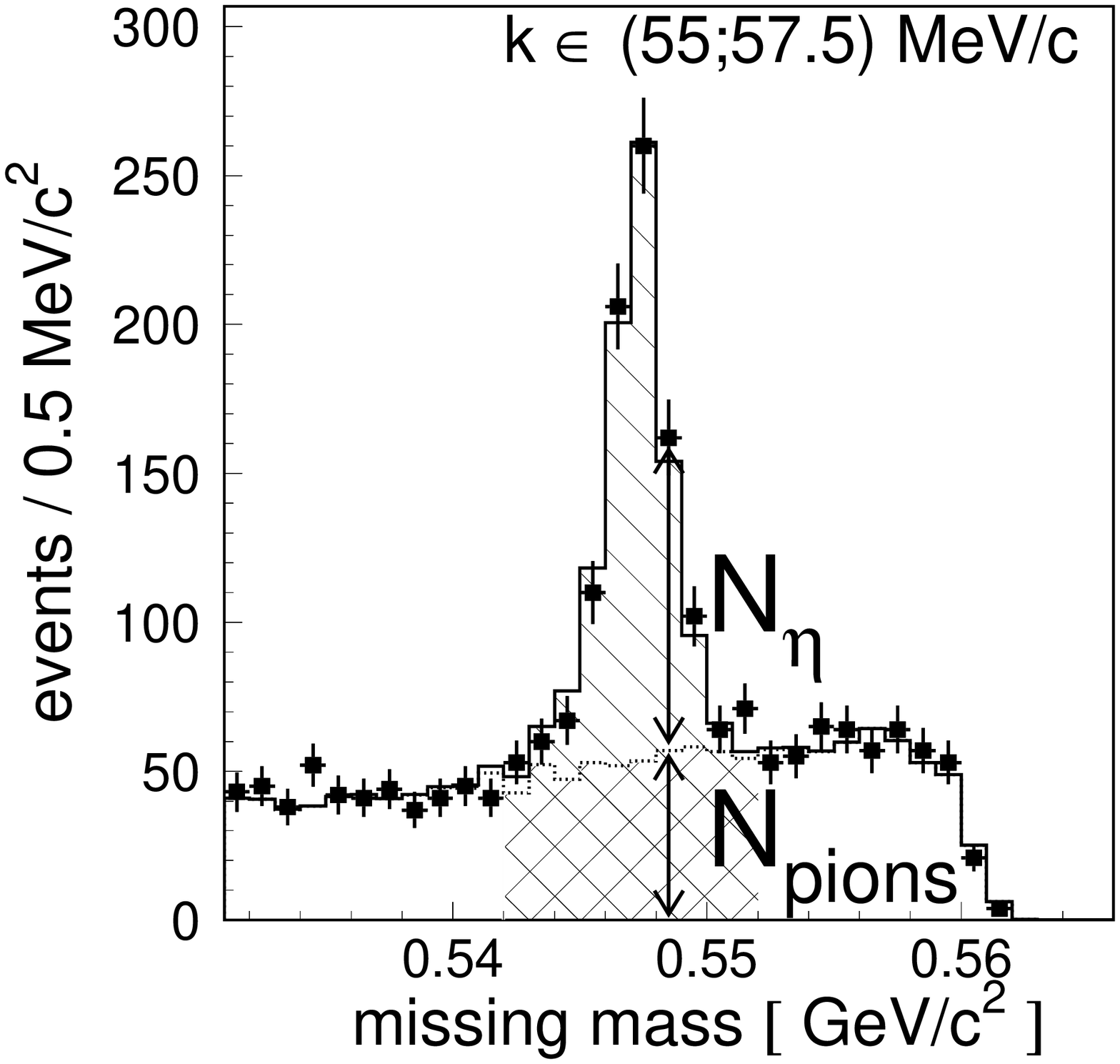}
\includegraphics[width=.32\textwidth]{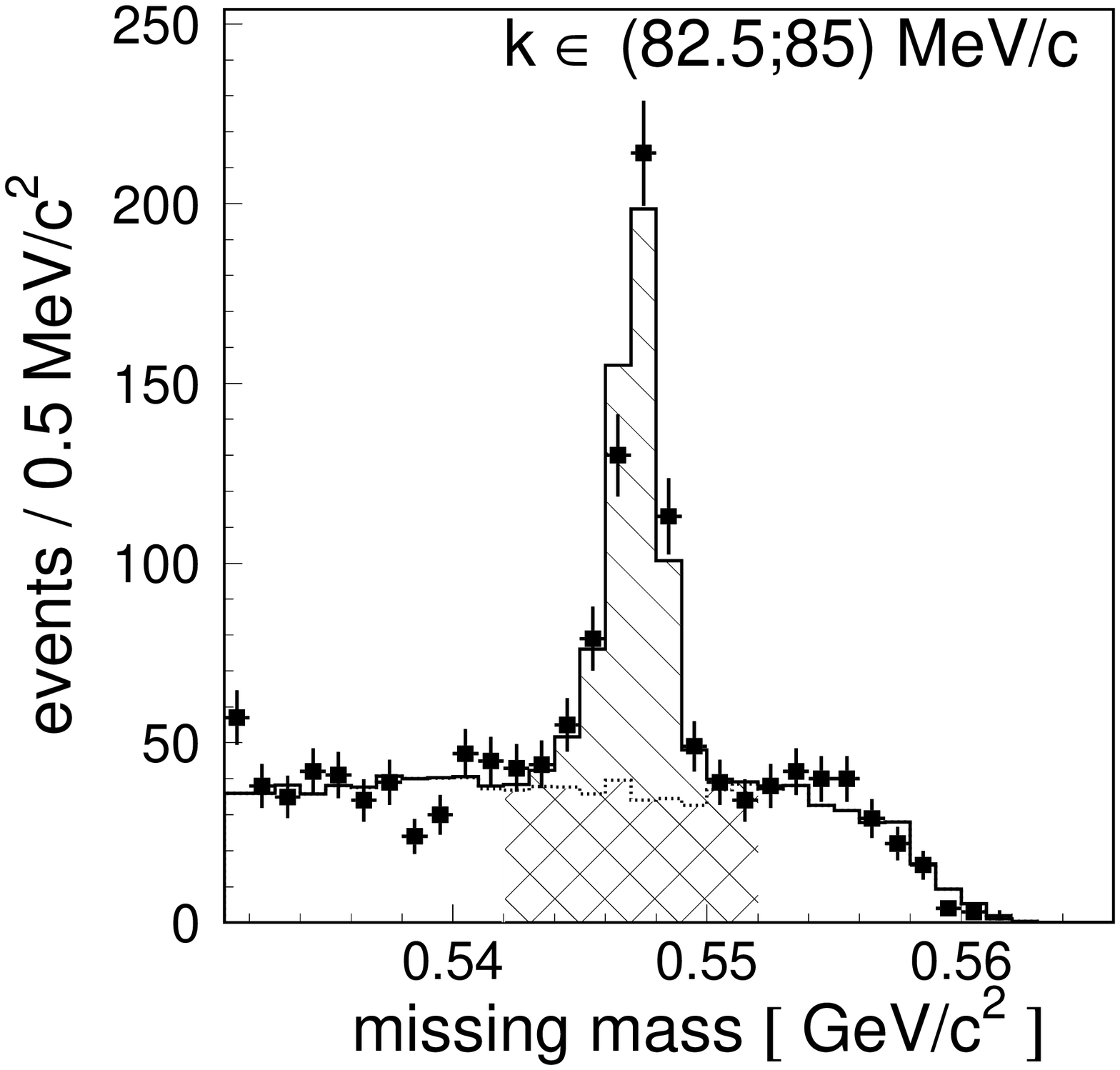}
\caption{
Examples of missing mass spectra measured~\cite{prc69} 
for the $pp \to ppX$
reaction at $k~\in~(27.5;~30.0)$~MeV/c, $k~\in~(55.0;~57.5)$~MeV/c
and $k~\in~(82.5;~85.0)$~MeV/c. 
Points represent experimental data. 
Dotted histograms depict simulations 
of the
$pp \to pp2\pi$, $pp \to pp3\pi$ and $pp \to pp4\pi$ 
reactions, while solid line histograms present the sum of 
simulations for the 
$pp \to pp\eta$ reaction and 
for the
$pp \to pp2\pi$, $pp \to pp3\pi$ and $pp \to pp4\pi$ reactions, respectively. The simulated
spectra were fitted to the data adjusting only the amplitude~\cite{prc69}.
\label{fig:stat_back}}
\end{figure}

Thus, $Y_{pp\eta}(k)$ can be extracted for each studied interval 
of $k$ by  grouping the sample of measured events according
to the value of $k$, next calculating 
the missing mass spectra of the $pp \to ppX$ reaction for each sub-sample separately,
and counting the number of $pp\eta$ events from these spectra.
Examples of typical missing mass histograms for three different $k$ intervals are presented 
in Figure~\ref{fig:stat_back}. 
The statistics obtained in the considered measurement allowed to divide  
the kinematically available range 
of $k$ into bins whose width ($\Delta{k}~=~2.5$~MeV/c)
corresponds approximately to the  accuracy of the 
determination of $k$ with ($\sigma(k)~\approx~2$~MeV/c). 
A clear signal of $\eta$ meson production is observed in each spectrum
on top of a continuous distribution originating from the multi-pion production.
It is important to note that in the studied range of the missing mass
the observed shape
of the multi-pion distributions is well reproduced in corresponding simulation studies for each region
of the phase space~\cite{habil}.
An extraction of $Y_{pp\eta}^*(k)$ - 
unbiased by the multi-pion production - is, however, not trivial. 
Applying a mixing technique one can   
construct the uncorrelated reference sample,
taking momentum vectors of protons corresponding to different real events.
A real event is determined by the momentum vectors of two protons registered in coincidence, 
and an uncorrelated event will thus comprise momentum vectors of protons ejected from different reactions.
Unfortunately, in such a sample of uncorrelated momentum vectors,
due to the loss of the kinematical bounds,
the production of the $\eta$ meson
will be not reflected on the missing 
mass spectrum and hence it cannot be used to extract a 
number of mixed-events corresponding to the production of the $\eta$ meson.
Therefore, in order to determine a background-free correlation function
for the $pp \to pp\eta$ reaction we performed the following analysis:
First, for each event, the probability $\omega$ was determined 
that this event corresponds to the $pp \to pp\eta$ reaction.
The probability $\omega_{i}$, that the $i^{th}~pp \to ppX$ event with a missing 
mass $m_{i}$, and a relative momentum of $k_{i}$ corresponds to a $pp \to pp\eta$
reaction was estimated according to the formula:
\begin{equation}
    \omega_{i} = \frac{N_{\eta}(m_{i},k_{i})}{N_{\eta}(m_{i},k_{i}) + N_{pions}(m_{i},k_{i})},
\label{equ:4}
\end{equation}
where $N_{\eta}$ stands for the number of 
the $pp \to pp\eta$ reactions and $N_{pions}$ 
is the number of events corresponding to the multi-pion
production with the invariant mass equal to $m_{i}$. 
The values of $N_{\eta}(m,k)$ and $N_{pions}(m,k)$ 
were extracted from the missing mass distributions produced 
separately for each of the studied intervals of $k$. 
An example of a missing mass spectrum 
with pictorial definitions of $N_{pions}$ 
and $N_{\eta}$ is presented in the middle panel of Figure~\ref{fig:stat_back}.
Now having introduced the weights we can calculate  also the value of $Y^*(k)$ 
separately for the $pp\eta$ and $pp(pions)$ final states.
We can achieve this by sorting an uncorrelated sample according to the $k$ values similarly as
in the case of the correlated events and next for each sub-sample 
we construct background free $Y^*_{pp\eta}(k)$ distributions as a sum of the probabilities that both protons in
an uncorrelated event originate from the reaction where the $\eta$ meson was created.
Specifically,  if in a given uncorrelated event denoted by $l$, one momentum is taken 
from a real event say $l1$
and the second momentum from another real event $l2$, then the probability that both correspond 
to reactions where an $\eta$ was created equals to $\omega_{l1} \cdot \omega_{l2}$,
and hence the uncorrelated yield $Y^*_{pp\eta}(k)$ may be constructed from the sum  
$\sum_l{\omega_{l1} \cdot \omega_{l2}}$,
where $l$ enumerates events in the uncorrelated 
sub-sample selected for a momentum range $k$.

Analogously one can calculate $Y^*_{pp(pions)}$ assigning to the event a weight equal to $1-\omega$.

\subsection{Acceptance corrections}
As the next necessary step in the data evaluation we corrected the determined yields
to account for the finite geometrical acceptance and detection efficiency
of the COSY-11 detectors~\cite{aip_klajus}. 
Hereafter for simplicity by "acceptance" both 
the geometrical acceptance and detection efficiency will be denoted.
The acceptance was calculated as a function of the proton momentum in the proton-proton rest frame $k$.
It was obtained using Monte-Carlo simulations and evaluated according to formula:
\begin{equation}
A(k) = \frac{N_{acc} (k)}{N_{gen} (k)},
\label{equ:6}
\end{equation}
where $N_{acc}(k)$ and $N_{gen}(k)$ denote the number of accepted and generated events, respectively.
Simulations were performed based on the GEANT-3 packages~\cite{geant}
including the realistic geometry of the detectors and a precise map
of the field of the dipole magnet. The momentum and spatial beam spreads, multiple scattering,
proton-proton final state interaction~\cite{swave_i_etap} 
and other known physical and instrumental effects were taken into account~\cite{brauksiepe,habil}.
Knowing the acceptance~\cite{aip_klajus}
it would be straightforward to correct the nominator
of equation~\ref{equ:3}, however the correction 
of the uncorrelated
yield $Y^*(k)$ is not trivial since the 
momenta of protons in the uncorrelated event originate from 
two idependent real events which in general could correspond to different
values of the detector acceptance.\\
Therefore, in order to derive a correlation function
corrected for the acceptance,  we have created a sample of data
that would have been measured with an ideal detector.
For this aim 
each experimental $pp\to pp\eta$ event was 
multiplied by 1/A(k).
This means that a given reconstructed
$pp\to pp\eta$ event with momentum of k was added to the experimental data sample $1/A(k)$ times.\\
Based on this corrected data sample  
we calculated the two-proton correlation function
according to equation~\ref{equ:3}. 
In order to avoid mixing between the same events, a   
"mixing step" in the calculations was set to a value bigger than 
the inverse of  the lowest acceptance value. 
The random repetition of identical combinations 
was also omitted by increasing correspondingly the "mixing step".  
In particular, a l$^{th}$ event, from the acceptance corrected data sample,
was "mixed" with a (l+n)$^{th}$
event, where $n~>~ max(1/A(k))$. 
If the (l+1)$^{th}$ event was the same as l$^{th}$,
then this was mixed with a (l+1+2n) event, etc.

\subsection{Results and conclusions}
\begin{figure}[h]
  \includegraphics[height=.27\textheight]{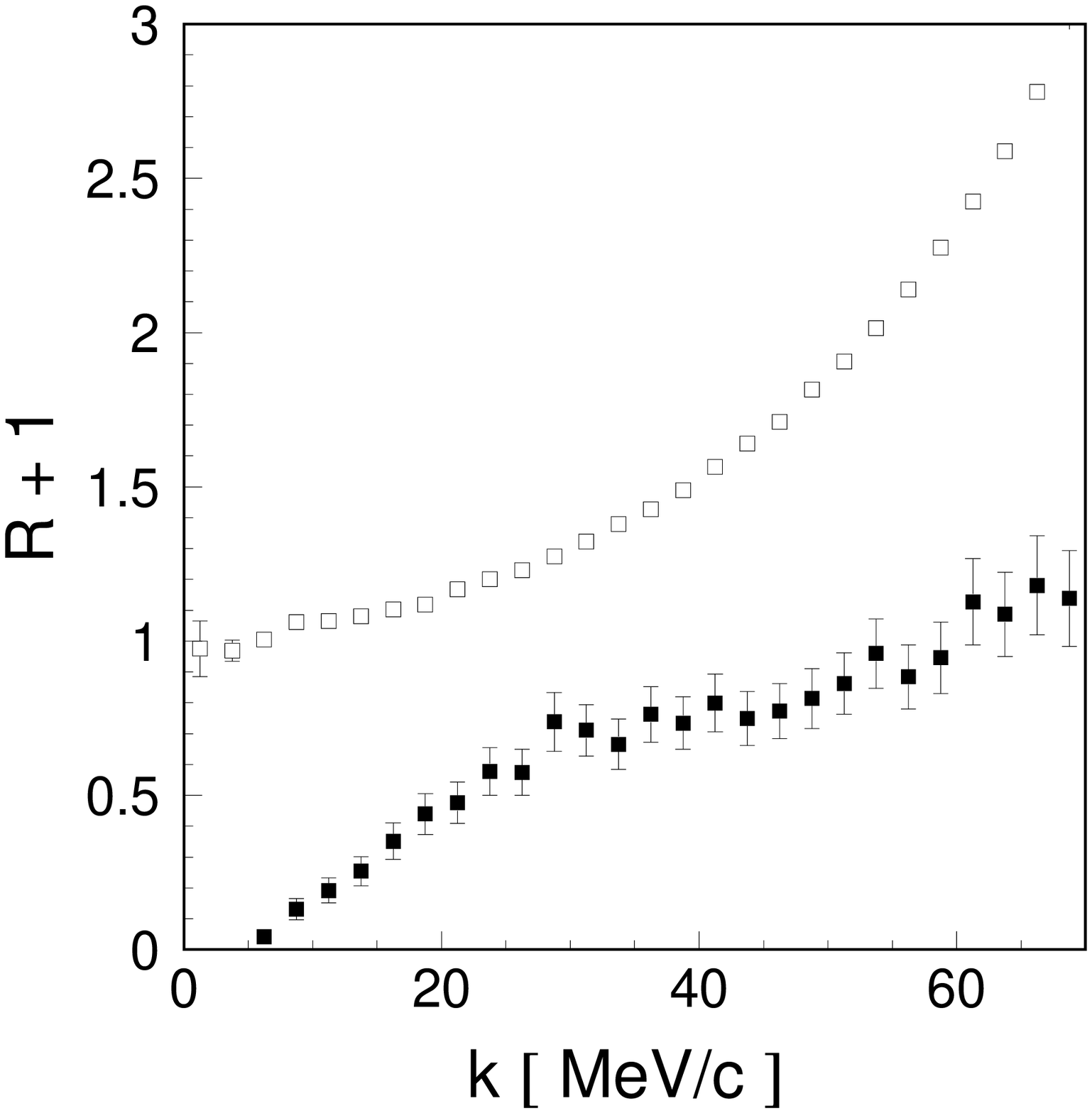} 
  \includegraphics[height=.27\textheight]{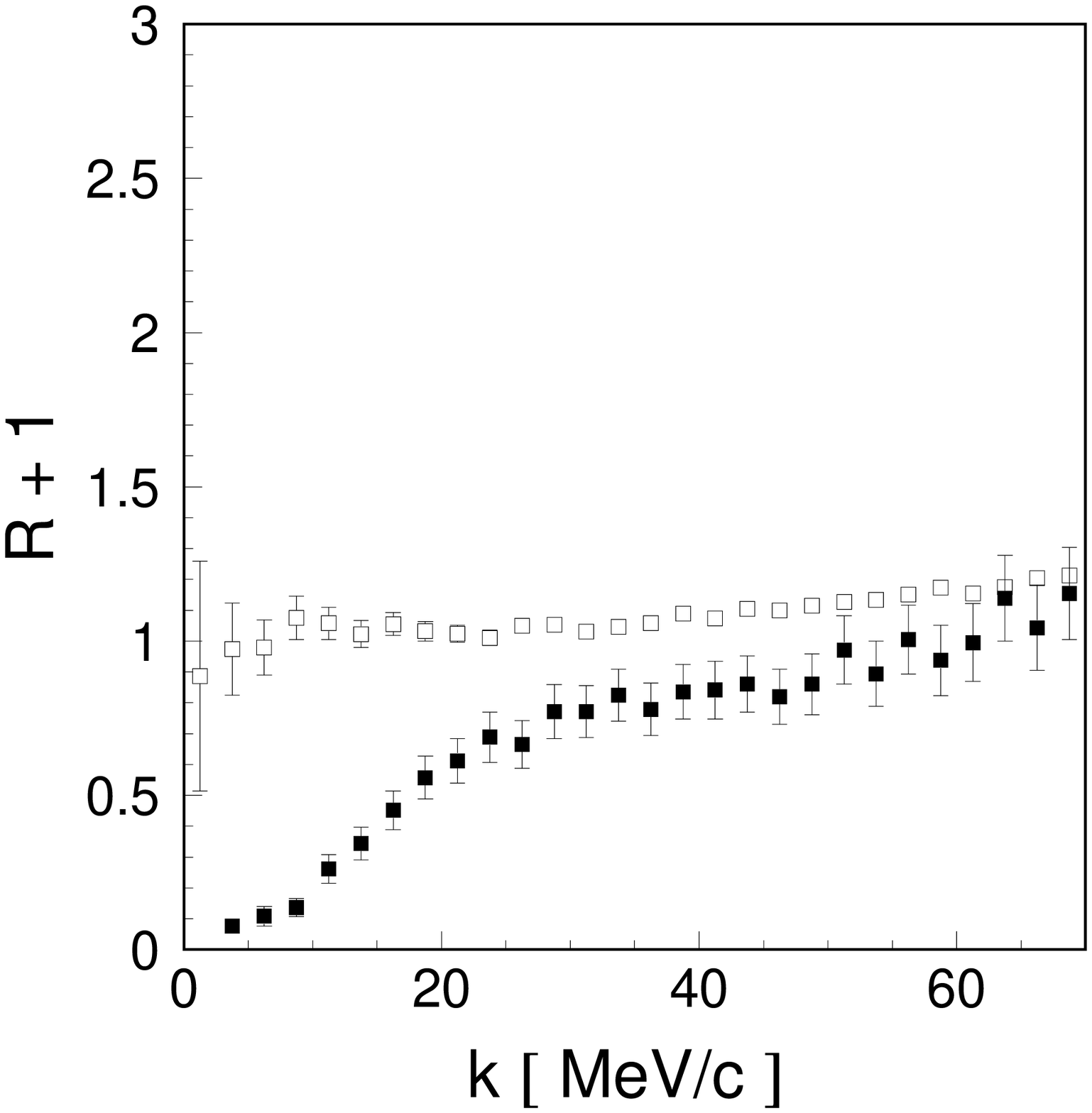}
  \caption{Acceptance corrected experimental proton-proton
 correlation functions for the production of the
  $\eta$ meson (left panel) 
  and multi-pions with the invariant mass equal to the mass of the $\eta$ meson (right panel).
  Full  squares denote the experimental data and open squares represent results
  of simulations determined for a point-like source assuming the homogeneous population 
  of the phase space for the reaction products.
\label{corr:mcexp}}
\end{figure}

The two-proton background-free correlation functions for the
$pp \to pp\eta$ and $pp \to pp+pions$
reactions corrected for the acceptances
are presented in  Figure~\ref{corr:mcexp} (full squares).
As mentioned in the introduction the shape of the obtained
correlation function reflects not only the
 space-time characteristics of the interaction volume
 but it may also be strongly modified
by the conservation of energy and momentum and by the final state interaction among the ejectiles.
In order to estimate the influence on the shape induced by the kinematical bounds 
we have constructed the correlation functions 
for both,  the $pp\to pp\eta$ and $pp\to pp+pions$ reaction
 assuming a point-like source using a Monte-Carlo simulation.
The  results of the simulations
are presented in Figure~\ref{corr:mcexp} (open squares) 
and it is apparent that they differ significantly 
from the experimental correlation function.
At the origin the Coulomb repulsion between protons
brings the experimental correlation function down to zero.

\begin{figure}
  \begin{center}
  \includegraphics[height=.45\textheight]{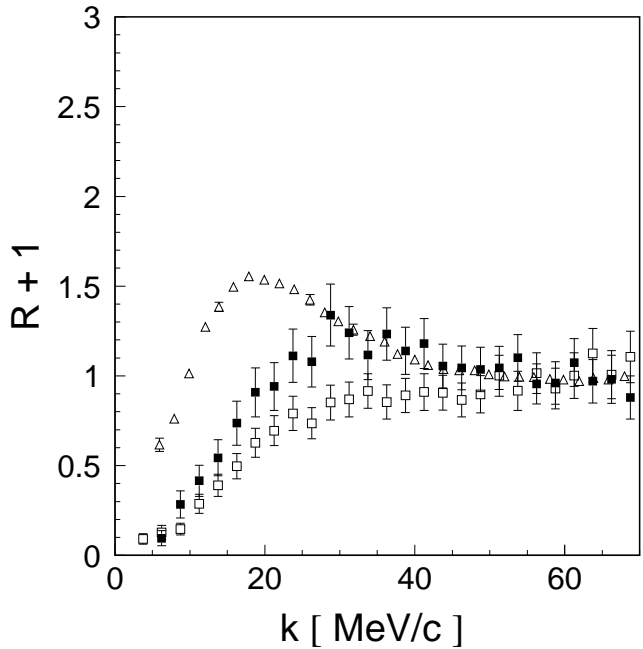}
  \end{center}
  \caption{The two-proton acceptance corrected
  correlation functions normalized to
  the corresponding simulated correlation function for a point-like source.
  Results  
  for the $pp \to pp\eta$ (full squares) and $pp\to pp+pions$ (open squares)
  are compared to the two-proton correlation function determined 
  from heavy ion collisions (triangles)~\cite{xx}.  
  \label{corr:double}}
\end{figure}

Next, in order to extract from the experimental data the shape 
of the correlation function free from the influence
of energy and momentum conservation we constructed a double ratio:
\begin{equation}
R(k)+1 = Const~\left(\frac{Y_{exp}(k)}{Y^*_{exp}(k)}~\slash~\frac{Y_{MC}(k)}{Y^*_{MC}(k)}\right),
\label{equ:5}
\end{equation}
where $Const$ denotes the normalization constant, and indices 'exp' and 'MC'
refer to the experimental
and simulated samples, respectively. 
The determined double ratios are given in Table~\ref{tabelka} and presented in Figure \ref{corr:double}.
The double ratio is a well established measure of correlations
used e.g. by the ALEPH, OPAL and DELPHI collaborations for studying the Bose-Einstein
or Fermi-Dirac correlations e.g. in the decays of Z boson~\cite{aleph1,delphi,opal} or W-pairs~\cite{aleph2}.
In Figure~\ref{corr:double} a significant discrepancy between the two-proton 
correlation functions determined from inclusive heavy ion reactions (triangles)~\cite{xx} 
and from exclusive proton-proton measurements (full squares) is clearly visible. The data from the kinematically
exclusive measurement do not reveal a peak structure at 20~MeV/c. 
Although a similar enhancement is also seen in the invariant mass of the proton-proton distributions~\cite{prc69}
it disappears in the correlation function for the $pp\to pp\eta $ and $pp\to pp+pions$ reactions.

At present it is not possible to draw a solid quantitative conclusion
about the size of the system since  e.g. in the case of the $pp\to pp\eta$ reaction
it would require to solve a three-body problem
where $pp$ and $p\eta$~\cite{wycech} interactions 
are not negligible and both contribute significantly to the proton-proton correlation. 
However, based on semi-quantitative predictions~\cite{deloff}
one can estimate that the system must be unexpectedly large 
with a radius in the order of 4~fm. This makes the result interesting in context 
of the predicted quasi-bound $\eta NN$ state \cite{ueda} and in view of the 
hypothesis~\cite{wycech_acta}  
that at threshold for the $pp\to pp\eta$ reaction 
the proton-proton pair may be emitted
from a large Borromean like object whose radius is about 4~fm.

\begin{table}[H]
\caption{\label{tabelka} The double ratios determined for the production of the
  $\eta$ meson and multi-pions with the invariant mass equal to the mass of the $\eta$ meson. 
  In the first column of the table $k$ denotes the center value of a 2.5 MeV/c interval.}
\begin{indented}
\item[]\begin{tabular}{@{}ccc}
\br
  k & $(R+1)_{\eta}$ & $(R+1)_{pions}$\\
  $[$MeV/c$]$ & &\\ 
\mr
   1.25 &  -----              &    -----\\
   3.75 &  -----              &    0.09 $\pm$  0.03\\
   6.25 &   0.10 $\pm$   0.04 &    0.13 $\pm$  0.04\\
   8.75 &   0.28 $\pm$   0.08 &    0.15 $\pm$  0.04\\
   11.25 &  0.42 $\pm$   0.09 &    0.29 $\pm$  0.05\\
   13.75 &  0.54 $\pm$   0.10 &    0.39 $\pm$  0.06\\
   16.25 &  0.74 $\pm$   0.12 &    0.50 $\pm$  0.07\\
   18.75 &  0.91 $\pm$   0.14 &    0.63 $\pm$  0.08\\
   21.25 &  0.94 $\pm$   0.13 &    0.70 $\pm$  0.08\\
   23.75 &  1.11 $\pm$   0.15 &    0.79 $\pm$  0.10\\
   26.25 &  1.08 $\pm$   0.14 &    0.74 $\pm$  0.09\\
   28.75 &  1.34 $\pm$   0.17 &    0.85 $\pm$  0.10\\
   31.25 &  1.24 $\pm$   0.15 &    0.87 $\pm$  0.10\\
   33.75 &  1.12 $\pm$   0.14 &    0.92 $\pm$  0.10\\
   36.25 &  1.23 $\pm$   0.15 &    0.86 $\pm$  0.09\\
   38.75 &  1.14 $\pm$   0.13 &    0.89 $\pm$  0.10\\
   41.25 &  1.18 $\pm$   0.14 &    0.91 $\pm$  0.10\\
   43.75 &  1.06 $\pm$   0.13 &    0.91 $\pm$  0.10\\
   46.25 &  1.04 $\pm$   0.12 &    0.87 $\pm$  0.10\\
   48.75 &  1.04 $\pm$   0.12 &    0.90 $\pm$  0.10\\
   51.25 &  1.04 $\pm$   0.12 &    1.00 $\pm$  0.11\\
   53.75 &  1.10 $\pm$   0.13 &    0.92 $\pm$  0.11\\
   56.25 &  0.95 $\pm$   0.11 &    1.02 $\pm$  0.11\\
   58.75 &  0.96 $\pm$   0.12 &    0.93 $\pm$  0.11\\
   61.25 &  1.07 $\pm$   0.13 &    1.00 $\pm$  0.13\\
   63.75 &  0.97 $\pm$   0.12 &    1.13 $\pm$  0.14\\
   66.25 &  0.98 $\pm$   0.13 &    1.01 $\pm$  0.13\\
   68.75 &  0.88 $\pm$   0.12 &    1.11 $\pm$  0.14\\
\br
\end{tabular}
\end{indented}
\end{table}

\vspace{0.3cm}
{\bf Acknowledgements:}\\
The work was partially supported by the
European Co\-mmu\-nity-Research Infrastructure Activity
under the FP6 and FP7 programmes (Hadron Physics,
RII3-CT-2004-506078, PrimeNet No. 227431), by
the Polish Ministry of Science and Higher Education under grants
No. 3240/H03/2006/31, 1202/DFG/2007/03, and 0084/B/H03/2008/34,
by the German Research Foundation (DFG),
and by the FFE grants from the Research Center J{\"u}lich, and by the virtual institute "Spin and strong QCD" 
(VH-VP-231).

\end{document}